\documentclass[prb,twocolumn,showpacs,amsmath,amssymb,superscriptaddress]{revtex4}

\usepackage{graphicx}
\usepackage{amssymb,amsmath}
\usepackage{dcolumn}
\usepackage{bm}
\usepackage{color}

\begin{document}

\title{Dispersion and polarization conversion of whispering gallery modes in arbitrary cross-section nanowires}

\author{G. Pavlovic}
\affiliation{LASMEA, Clermont Universit\'{e}–Université Blaise Pascal, BP 10448, 63000
Clermont-Ferrand, France}
\affiliation{LASMEA, UMR 6602, CNRS, 63177 Aubi\'{e}re, France}
\author{G. Malpuech}
\affiliation{LASMEA, Clermont Universit\'{e}–Université Blaise Pascal, BP 10448, 63000
Clermont-Ferrand, France}
\affiliation{LASMEA, UMR 6602, CNRS, 63177 Aubi\'{e}re, France}
\author{N. A. Gippius}
\affiliation{LASMEA, Clermont Universit\'{e}–Université Blaise Pascal, BP 10448, 63000
Clermont-Ferrand, France}
\affiliation{LASMEA, UMR 6602, CNRS, 63177 Aubi\'{e}re, France}
\affiliation{A. M. Prokhorov General Physics Institute, Russian Academy of Sciences,
Vavilova Street 38, Moscow 119991, Russia}

\date{\today}

\begin{abstract}
We investigate theoretically the optical properties of Nano-Wires (NWs) with cross sections having either discrete or cylindrical symmetry. The material forming the wire is birefringent, showing a different dielectric response in the plane and along the axis of the wire, which is typically the case for wires made of wurtzite materials, such as ZnO or GaN. We look for solutions of Maxwell`s equations having the proper symmetry. The dispersions and the linewidths versus angle of incident light for the modes having high momentum in the cross-section plane, so called whispering gallery modes, are calculated. We put a special emphasis on the case of hexagonal cross sections. The energy positions of the modes for a set of  azimuthal quantum numbers are shown. We demonstrate the dependence of the energy splitting between TE and TM modes versus birefringence.
The polarization conversion from TE to TM with increase of the axial wave vector
is discussed for both cylindrical and discrete symmetry.

\end{abstract}

\maketitle

\section{Introduction}

Nano-Wires are the objects with a cross-section dimension $a$ reduced to nanoscale values, several orders of magnitude smaller than their length $L$. The dynamics along the wire axis can be decoupled from the transverse one in case $L>>a$ and the system can be treated as quasi one-dimensional. Low dimensional systems in general reveal novel phenomena and are excellent candidates for applications in new technologies. NWs could be used in communications \cite{Motayed}, quantum computation \cite{Schenkel}, or biological sensors \cite{Hahm}.

In the last years, significant progress has been achieved in the growth of semiconductor NWs. Many new interesting optical effects have been reported for such structures, like ultraviolet lasing under optical pumping in ZnO NWs \cite{Huang} and recently, polaritonic effects. 1D exciton-polaritons \cite{Cavity} appear due to the strong coupling of excitons with photonic whispering gallery modes (WGMs) \cite{Trichet}. These modes owe their name to their similarity with acoustic resonances in real galleries. They propagate in the NW's cross-section and due to azimuthal momentum undergo multiple internal reflections. The number of these reflections can be very high, resulting in a high quality factor.

The problem of resonant mode frequencies and of their life time in dielectric resonators has been studied for the cylindrical and hexagonal cross-sections both in mesoscopic (large $ka$) \cite{Wiersig} and microscopic (small $ka$) \cite{Nobis} regimes; where $k$ is the wave vector of the incident light. NWs with cylindrical cross-section are very well theoretically studied in the isotropic case, being a textbook subject \cite{Jackson}. On the other hand, discrete cross-sections symmetries are much less studied whereas they are of particular interest because real structures often have polygonal cross-sections. Wurtzite or diamond crystals (like ZnO or GaN) generally form hexagonal NWs, but rectangular and triangular shaped NWs are also possible \cite{Yang}.

A remaining theoretical task in these discrete symmetry cross section systems is to calculate the dispersion of WGMs (dependence of their energy versus the axial wave vector), as well as their linewidth. The impact of the anisotropy of the dielectric response of wurtzite materials on the polarization eigenstates of the WGMs has also not been addressed so far, to the best of our knowledge. On the other hand, it is well known that semiconductors with wurtzite structure possess birefringent optical anisotropy \cite{Wang}. NWs fabricated from such materials have directional dielectric response characterized by two refractive indexes - along the axis of anisotropy $n_{z}$ and perpendicular to it $n_{r} $. Another source of birefringence $\delta n=n_{z}-n_{r}$ comes from axial variation of the NW radius due to inevitable growth imperfections.

In this paper, we determine the optical eigenmodes of NWs of arbitrary cross-sections, considering both isotropic and anisotropic media. In the case of circular NW cross section, the solutions are cylindrical harmonics of particular azimuthal quantum number $m$. For arbitrary discrete geometries, the appropriate linear combination of cylindrical harmonics is used to fulfill the boundary conditions for the tangential electric and magnetic fields on the NW surface. The dispersions, line-widths, polarization, and field distribution of the modes versus the angle of incident light (longitudinal wave vector) are deduced from this model. In case of birefringent media, the $k_{z} =0$ WGMs are either TE or TM polarized. For other longitudinal wave vectors, strong mixing of polarization occurs in both isotropic and anisotropic NWs leading to the formation of hybrid modes - EH and HE, depending on whether electric or magnetic field dominates in z-direction. Nevertheless, in this paper we will use the notation TE for the former and TM for the later.

\section{System and mode symmetries}

 \begin{figure}[fbp]
 \includegraphics[width=0.99\linewidth]{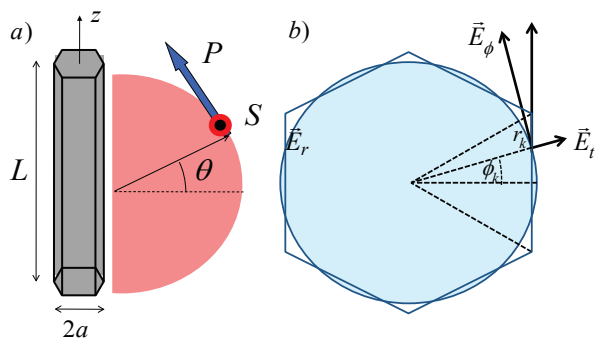}
 \caption{$a)$ Hexagonal cylinder of finite height and the radius of its circum circle. The arrows show the plane of $S$ and $P$ polarized light. $\theta$ is the angle of incidence $b)$ Tangential component of the electric field on the surface of the hexagon at the point $(r_{k},\phi_{k})$}
 \label{fig1}
 \end{figure}

We consider an infinitely long wire with its main axis along the z-direction and having either a circular or a polygonal cross section. The Figure~\ref{fig1}(a) shows a finite length hexagonal cross section wire. The translational symmetry along the z-direction imposes an additional factor in the wave functions having the phase form $e^{ik_{z} z}$. $k_{z} $ is a good quantum number which can take any real value for an infinitely long wire.

The axial symmetry of regular polygons is the $n$-fold rotation $C_{n}$, where $n$ is the number of polygon sides. For a circular cross section $n=\infty$, which corresponds to rotation by infinitesimally small angles around the z-axis. The symmetry group $C_{\infty }$ is continuous and, like in the case of translation, there is a good quantum number $m\in {\mathbb Z}$. Then, the $\varphi $-dependence of eigenmodes for each $m$ is given by a single angular harmonic $e^{im\varphi}$.

For regular polygons, $n$ is finite: triangle $C_{3}$, square $C_{4}$, pentagon $C_{5}$, hexagon $C_{6}$ etc. These systems remain invariant for in-plane rotations by corresponding discrete angles $2\pi m/n$.  The existence of the minimal angle of rotation in these symmetries results in finite number of irreducible representations of the group with $m=0,\pm 1,...,\pm (n-2)/2,n/2$ ($n$ even) or $m=0,\pm 1,...,\pm (n-1)/2$ ($n$ odd) \cite{Jones}. The transformation properties of higher $m$-s appear to be equivalent to those of the first Brillouin's zone in $m$-space. The eigenmodes behavior along $\varphi$ coordinate is no more given by a single angular harmonic, but rather by their linear combination.

Another important class of polygonal cross section symmetries are reflections - vertical mirror symmetries.  The hexagon, for example, is unchanged under reflections in the six vertical planes: 3 containing big hexagon diagonals and 3 connecting the centers of the opposite sides. These operations $\sigma _{d,i}$ and $\sigma _{v,i}$ transform in-plane rotations by the angle $\varphi$,$R(\varphi )$, into rotations by the opposite angles -$\varphi$, $R(-\varphi )$: $\sigma _{v,d} R(\varphi )\sigma _{v,d} =R(-\varphi )$. An important physical consequence is the equivalence of $m$ and $-m$. The whole symmetry group describing the axial symmetry of polygonal cross-section NWs containing both $n$-fold rotations and mirror reflections is denoted with $C_{nv}$.

The electric $\vec{E}=(E_{r},E_{\varphi },E_{z})$ and magnetic field $\vec{H}=(H_{r},H_{\varphi },H_{z})$ transform differently under reflections. The electric field is a polar (ordinary) vector and magnetic field, being the cross-product $\vec{H}=(1/\mu )\vec{\nabla }\times \vec{A}$, is a pseudovector.  Reflections change an eigenmode $(E_{r},H_{r},E_{\varphi },H_{\varphi },E_{z},H_{z})$ to  $(E_{r},-H_{r},-E_{\varphi},H_{\varphi},E_{z},-H_{z})$ . It is interesting to see how the usual transverse electric (TE) and transverse magnetic (TM) eigenmodes for $k_{z} =0$ are modified under reflection.
TE modes transform from $(E_{r},E_{\varphi},H_{z})$ to $(E_{r},-E_{\varphi},-H_{z})$, i.e. their parity is opposite to that of the scalar function $H_{z}$.
TM modes transform from $(H_{r},H_{\varphi},E_{z})$ to $(-H_{r},H_{\varphi},E_{z})$ after reflection, preserving the parity of the scalar function $E_{z}$.

\section{Formalism}

 \begin{figure}[fbp]
 \includegraphics[width=0.99\linewidth]{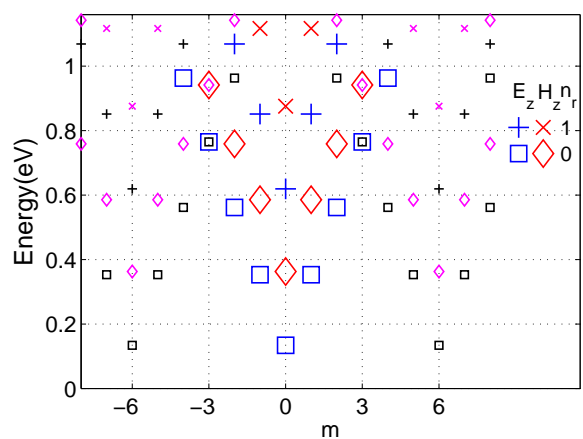}
 \caption{ Modes of a cylindrical cross-section nanowires versus angular momentum $m$ (large symbols). The small symbols represent the modes of an hexagonal cross section nanowire which are coming form the superposition of the modes of the cylinder having an hexagonal symetry.}
 \label{fig2}
 \end{figure}

 \begin{figure}[fbp]
 \includegraphics[width=0.99\linewidth]{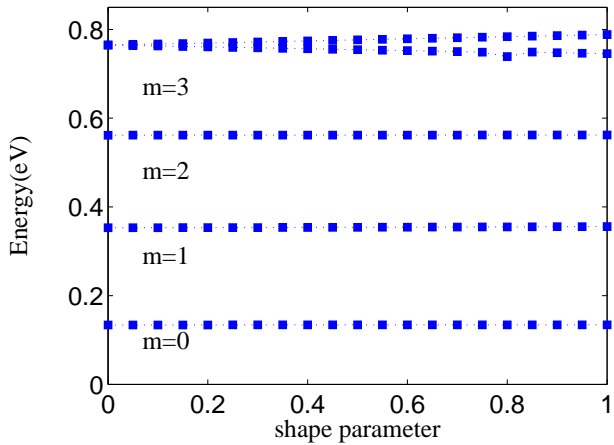}
 \caption{ Cylindrical mode evolution versus the shape parameter $x$ which varies from 0 for a cylinder to 1 for an hexagonal wire}
 \label{fig3}
 \end{figure}

We start with the Maxwell's equations written in the frequency domain

\begin{equation} \label{1}
\nabla \times \vec{H}(\vec{r},\omega )=-i\frac{\omega }{c} \vec{D}(\vec{r},\omega )
\end{equation}

\begin{equation} \label{2}
\nabla \times \vec{E}(\vec{r},\omega )=i\frac{\omega }{c} \vec{B}(\vec{r},\omega )
\end{equation}

\begin{equation} \label{3}
\nabla \vec{D}(\vec{r},\omega )=0
\end{equation}

\begin{equation} \label{4}
\nabla \vec{B}(\vec{r},\omega )=0
\end{equation}

Taking the curl of the second equation and using  $\vec{B}(\vec{r},\omega )=\mu \vec{H}(\vec{r},\omega )$ from Eq.(\ref{1}) follows the wave equation

\begin{equation} \label{5}
\nabla \times \nabla \times \vec{E}(\vec{r},\omega )-\frac{\omega ^{2} }{c^{2} } \mu \vec{D}(\vec{r},\omega )=0.
\end{equation}

Displacement field is given by $\vec{D}(\vec{r},\omega )=\varepsilon \vec{E}(\vec{r},\omega )$, and we consider anisotropic permittivity $\varepsilon $ of the form $\varepsilon=diag(\varepsilon _{r},\varepsilon _{r},\varepsilon _{z} )$, where $\varepsilon _{r}$ and $\varepsilon _{z}$ are the permittivities in the NW cross section and along the z-axis respectively.

The double curl of the electric field in the equation (\ref{5}) can be rewritten as

\begin{equation} \label{6}
\nabla \times \nabla \times \vec{E}=\nabla (\nabla \vec{E})-\Delta \vec{E}.
\end{equation}

In an isotropic medium this equation immediately gives $\nabla \vec{E}=0$ and Eq.(\ref{6}) reduces to $-\Delta \vec{E}$. In an anisotropic medium, both terms of the equation (\ref{6}) are in general non-zero. However, one still can decompose the waves into pure TE ones (for which $E_{z} =0$ and $\nabla \vec{E}=\vec{k}\vec{E}=0$) and pure TM ones (for which $H_{z} =0$ and $\nabla \vec{E}=\vec{k}\vec{E}\ne 0$).

Let's now derive the equation for $E_{z}$ for the TM wave from equation (\ref{6}). Using the equation (\ref{3}) we calculate $\nabla \vec{E}={\partial E_{z} \mathord{\left/{\vphantom{\partial E_{z}\partial z}} \right.\kern-\nulldelimiterspace}\partial z}\left(1-{\varepsilon _{z}\mathord{\left/{\vphantom {\varepsilon _{z}\varepsilon _{r}}}\right. \kern-\nulldelimiterspace}\varepsilon _{r}}\right)$ which is non zero in case of TM waves in anisotropic materials. The $z$-component of equation (\ref{6}) reads

\begin{equation} \label{7}
\left.\nabla(\nabla\vec{E})\right|_{z}=\frac{\partial^{2} E_{z}}{\partial z^{2}}\left(1-\frac{\varepsilon _{z} }{\varepsilon _{r} }\right).   \end{equation}

Inserting the equation (\ref{7}) in the wave equation (\ref{5}) allows obtaining the equation for the $E_{z}$ for the TM wave:

\begin{equation} \label{8}
\frac{\partial^{2} E_{z} }{\partial z^{2} }\left(\frac{\varepsilon _{z} }{\varepsilon _{r}} -1\right)+\Delta E_{z} +\frac{\omega ^{2} }{c^{2} }\varepsilon _{z} E_{z}=0.
\end{equation}

The equation for the $H_{z}$ field for the TE wave has the form of the usual Helmholtz's equation for an isotropic problem:

\begin{equation} \label{9}
\Delta H_{z}+\frac{\omega ^{2} }{c^{2} }\varepsilon_{r}H_{z}=0.
\end{equation}

So far we have considered infinite homogeneous media. The solution for the wires can be obtained by matching the boundary conditions for the bulk waves on the wire surface. These boundary conditions mix bulk TE and TM waves, making decomposition into pure TE and TM modes only possible for $k_{z} =0$. Because of the non-separability of TE and TM modes, we have to simultaneously solve equations (\ref{8}) and (\ref{9}).
For cylindrical wires, we search for the solution in the form $R_{E,H}(r)e^{i(m\varphi +k_{z} z-\omega t)}$ as it follows from the section on the mode symmetry. The radial parts $R_{E,H}(r)$ satisfy

\begin{equation} \label{10}
\frac{d}{dr} \left(r\frac{dR_{E,H} (r)}{dr} \right)+\left(rq_{E,H}^{2} -\frac{m^{2} }{r} \right)R_{E,H} (r)=0,
\end{equation}

where

\begin{equation} \label{11}
q_{H} =q_{E} (\varepsilon _{r} /\varepsilon _{z} )^{1/2} =(\varepsilon _{r} \omega ^{2} /c^{2} -k_{z}^{2} )^{1/2}
\end{equation}
are in plane wave vectors, of magnetic and electric field within the wire. In the isotropic case $q_{H} =q_{E}$. Outside the NW, the same equation (\ref{10}) holds for the radial part with $q_{out} =(\varepsilon _{out} \omega ^{2} /c^{2} -k_{z}^{2} )^{1/2} $; $\varepsilon _{out} $ is the dielectric constant of wire's environment that will be assumed to be isotropic.

The equation (\ref{10}) is a Bessel's differential equation. The solutions inside the wire are the linear combinations of Bessel's functions of the first kind $J_{m} (x)$, whereas propagating solutions outside are the linear combinations of Hankel's functions $H_{m}^{1} (x)$ of the first kind. Therefore one can look for the solution for $E_{z}^{in}$ and $H_{z}^{in} $ fields in the following form:

\begin{equation} \label{12}
E_{z}^{in}=\sum_{m}A_{m}J_{m}(q_{E} r)\phi_{m};
H_{z}^{in}=\sum_{m}B_{m}J_{m}(q_{H} r)\phi_{m}
\end{equation}

\begin{equation} \label{13}
E_{z}^{out}=\sum _{m}C_{m}H_{m}^{1} (q_{out} r)\phi _{m};
H_{z}^{out}=\sum _{m}D_{m}H_{m}^{1} (q_{out} r)\phi _{m}
\end{equation}
where $\phi _{m} =\exp (i(m\varphi +k_{z} z-\omega t))$. The transverse in-plane components of the fields can be deduced from the z-components (12,13) through the following equations:
\begin{widetext}
\begin{equation} \label{14}
q_{}^{2} \vec{E}_{T} =i\frac{\omega }{c} \nabla _{T} \times \vec{H}_{z} +ik_{z} \nabla _{T} \vec{E}_{z},
\end{equation}

\begin{equation} \label{15}
q_{}^{2} \vec{H}_{T} =-i\frac{\omega }{c} \nabla _{T} \times \vec{E}_{z} +ik_{z} \nabla _{T} \vec{H}_{z},
\end{equation}

where $q=q_{H,E}$ for internal fields and $q=q_{out}$ for external field. Inside the wire we directly find:

\begin{equation} \label{16}
E_{\varphi }^{in} =-\sum _{m}\left(A_{m} \frac{mk_{z} }{q_{H}^{2} r} J_{m} (q_{E} r)+iB_{m} \frac{\omega }{cq_{H}^{} } J_{m}^{'} (q_{H} r)\right)\phi _{m},
\end{equation}

\begin{equation} \label{17}
E_{r}^{in} =\sum _{m}\left(iA_{m} \frac{k_{z} q_{E}^{} }{q_{H}^{2} } J_{m}^{'} (q_{E} r)-B_{m} \frac{m\omega }{cq_{H}^{2} r} J_{m} (q_{H} r)\right) \phi _{m},
\end{equation}

\begin{equation} \label{18}
H_{\varphi }^{in} =\sum _{m}\left(iA_{m} \frac{\omega \varepsilon _{z} }{cq_{E}^{} } J_{m}^{'} (q_{E} r)-B_{m} \frac{mk_{z} }{q_{H}^{2} r} J_{m}^{} (q_{H} r)\right)\phi _{m},
\end{equation}

\begin{equation} \label{19}
H_{r}^{in} =\sum _{m}\left(A_{m} \frac{m\omega \varepsilon _{r} }{cq_{H}^{2} r} J_{m} (q_{E} r)+iB_{m} \frac{ik_{z} }{cq_{H}^{} } J_{m}^{'} (q_{H} r)\right)\phi _{m}.
\end{equation}

The outside fields read:

\begin{equation} \label{20}
E_{\varphi }^{out} =-\sum _{m}\left(C_{m} \frac{mk_{z} }{q_{out}^{2} r} H_{m}^{1} (q_{out} r)+iD_{m} \frac{\omega }{cq_{out} } H_{m}^{1'} (q_{out} r)\right)\phi _{m},
\end{equation}

\begin{equation} \label{21}
E_{r}^{out} =\sum _{m}\left(C_{m} \frac{ik_{z} }{q_{out}^{} } H_{m}^{1'} (q_{out} r)-iD_{m} \frac{m\omega }{cq_{out}^{2} r} H_{m}^{1} (q_{out} r)\right)\phi _{m},
\end{equation}

\begin{equation} \label{22}
H_{\varphi }^{out} =\sum _{m}\left(C_{m} \frac{\omega \varepsilon _{out} }{cq_{out}^{} } H_{m}^{1'} (q_{out} r)-iD_{m} \frac{mk_{z} }{q_{out}^{2} r} H_{m}^{1} (q_{out} r)\right) \phi _{m},
\end{equation}

\begin{equation} \label{23}
H_{r}^{out} =\sum _{m}\left(C_{m} \frac{m\omega \varepsilon _{out} }{cq_{out}^{2} r} J_{m} (q_{out} r)+iD_{m} \frac{k_{z} }{q_{out}^{} } J_{m}^{'} (q_{out} r)\right) \phi _{m}.
\end{equation}
\end{widetext}

From these expressions, one can check that it is not possible to have a pure TE mode, putting $A_{m} =0$ (no longitudinal electric field $E_{z} =0$). For $k_{z} \ne 0$, the magnetic field is also present in the transverse plane and the last two terms in the equations (\ref{18},\ref{19}) are non-zero. Similar argumentation holds for the existence of a pure TM mode. In the general case of $k_{z} \ne 0$, a sharp separation of TE and TM modes is no more possible and the modes are mixed.

To calculate the mode dispersion, one needs to impose the proper boundary conditions for both electric and magnetic fields resulting in the system of equations for the fields amplitudes. The tangential external and internal electric and magnetic fields should match on the wire boundary. If we denote $F_{in,out} =(E_{z}^{in,out} ,E_{t}^{in,out} ,H_{z}^{in,out} ,H_{t}^{in,out} )^{T}$, boundary conditions are fulfilled if $\delta F=F_{in} -F_{out}$ is zero on the surface of the wire. In the case of cylindrical wires of radius $a$, the fields for a single angular harmonics are proper solutions, and the last statement is valid for each of them independently. Tangential components on the boundary are $\varphi $-field (equations \ref{16} and \ref{18}) and for a chosen $m$, on the boundary $r=a$ we can write matrix equation $F_{m} (\omega ,k_{z} ;a)X_{m} =0$. The vector $X_{m}$ being the set of partial amplitudes  $X_{m} =(A_{m} ,B_{m} ,C_{m} ,D_{m} )$ and matrix $F_{m} (\omega ,k_{z} ;a)$ given by the following expression:

\begin{widetext}
\begin{equation} \label{24}
F_{m} (\omega ,k_{z} ;a)=\left[\begin{array}{cccc} {J_{m} (q_{E} a)} & {0} & {-H_{m}^{1} (q_{out} a)} & {0} \\ {-\frac{mk_{z} }{q_{H}^{2} a} J_{m} (q_{E} a)} & {-\frac{i\omega }{cq_{H}^{} a} J_{m}^{'} (q_{H} a)} & {\frac{mk_{z} }{q_{out}^{2} a} H_{m}^{1} (q_{out} a)} & {\frac{i\omega }{cq_{out}^{} } H_{m}^{1'} (q_{out} a)} \\ {0} & {J_{m} (q_{H} a)} & {0} & {-H_{m}^{1} (q_{out} a)} \\ {\frac{i\omega \varepsilon _{z} }{cq_{E}^{} } J_{m}^{'} (q_{E} a)} & {-\frac{mk_{z} }{q_{H}^{2} a} J_{m} (q_{H} a)} & {\frac{i\omega \varepsilon _{z} }{cq_{out}^{} } H_{m}^{1'} (q_{out} a)} & {\frac{mk_{z} }{q_{out}^{2} a} H_{m}^{1} (q_{out} a)} \end{array}\right].
\end{equation}
\end{widetext}

Dispersion relation $\omega _{m} (k_{z} )$ is calculated from the condition that the determinant of a homogeneous system of equations must be zero in order to have a nontrivial solution $X_{m} \ne 0$.  These dispersions can be found by efficient iterative numerical procedure \cite{Tikhodeev}, based on the linearization of the matrix $F_{m} (\omega ,k_{z} ;a)$ with respect to $\omega$.

The Figure~\ref{fig2} shows ( large simbols) the calculated mode energies at  $k_{z}=0$ for a cylindrical NW. For each angular momentum $m$, and for each polarization $(E_{z} , H_{z} )$ a series of modes characterized by their radial quantum number $n_{r}$  are showing up. For simplicity we show only the modes $n_{r}=0$ and $n_{r}=1$ on the Figure~\ref{fig2} calculated for a radius $a=0.6\mu m$ and index of refraction $n=2.52$.

Before discussing the details of the calculation procedure for non-cylindrical NWs, it is helpful to consider qualitatively the expected modification of the mode spectra induced by the deviation of the NW shape from the cylindrical geometry. When NW cross section is just slightly deviated from a circle, the qualitative description of the modes can be obtained by perturbative considerations. The small deviation of the boundary cross-section results in the mixing between the cylindrical modes of different momenta. The physical reason of this mixing is the additional scattering of a cylindrical mode on a non-cylindrical boundary. The 'selection rules' of this scattering obviously depend on the symmetry of the NW. In case of an hexagonal perturbation e.g., that is invariant in respect to any rotation on multiples of $2\pi/6$ , the changes of angular momentum are the multiples of 6. As a result, each cylindrical mode acquires a 'tail' of other harmonics whose amplitude depends on the energy mismatch between the energy of the main mode and that of the member of the 'tail'. The less is the energy mismatch, the stronger is the admixture of the other state to the initial cylindrical mode.
The states with opposite $m$ are degenerate in the absence of magnetic field and consequently can be strongly coupled by the surface perturbation, in case the angular momentum difference corresponds to that of boundary perturbation.

To illustrate the mixing between the modes the Figure~\ref{fig2} shows the structure of the cylindrical modes for small angular momenta with superposed harmonics with the periodicity of a hexagon.

For the hexagon, a strong coupling of degenerate harmonics occurs for $m-(-m)=6k$, and results in a strong mixing of the modes with e.g. $m=\pm 3$
. As a result, the  $m=\pm 3$ doublet splits into a pair of singlet states. The dependence of this splitting on the hexagonal perturbation is illustrated on the Figure~\ref{fig3} for a set of cross-sections smoothly varying from a cylinder (parameters of \ref{fig2}) to an hexagon with the same area of the cross-section. The radial position of the boundary for each angle was obtained by linear interpolation between the cylinder ($x=0$) and the hexagon ($x=1$), where $x$ is the shape parameter.

Let us now discuss in more details the calculation procedure for non-cylindrical NWs. We assume that the solutions inside and outside the wire can be written as linear combinations of the cylindrical harmonics. Boundary conditions in this case are more complicated. Single angular harmonics are no more proper solutions because they cannot match the tangential field boundary conditions on a boundary changing with the azimuth. The solution has to be written as an infinite combination of cylindrical harmonics as shown in the equations (\ref{16}-\ref{23}). Moreover, because the normal to the boundary is not parallel to the radial vector, the tangential fields on the wire's surface contain both radial and azimuthal components ( Figure~\ref{fig1}(b)):

\begin{eqnarray} \label{25}
\vec{E}_{t}^{in,out} (r_{k} ,\varphi _{k} )=E_{r}^{in,out} (r_{k} ,\varphi _{k} )\sin (\vec{\tau }\vec{e}_{r} )\vec{e}_{r} +\\
E_{\varphi }^{in,out} (r_{k} ,\varphi _{k} )\cos (\vec{\tau }\vec{e}_{\varphi } )\vec{e}_{\varphi }. \nonumber
\end{eqnarray}

\begin{figure}[fbp]
 \includegraphics[width=0.99\linewidth]{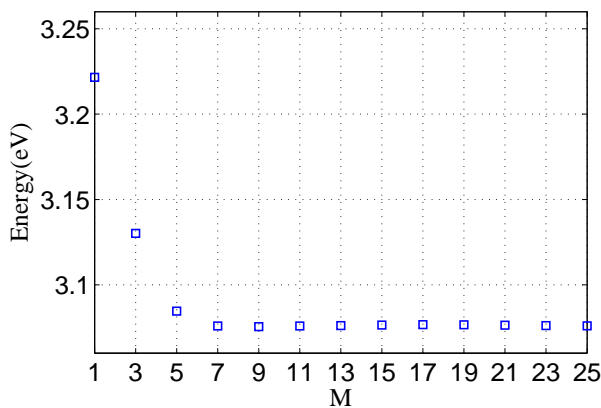}
 \caption{ Convergence of the mode $m=0$ being in the energy range 3.05-3.25 eV versus the number of cylindrical harmonics $M$ used in the calculation}
 \label{convergence}
 \end{figure}

 \begin{figure}[fbp]
 \includegraphics[width=0.99\linewidth]{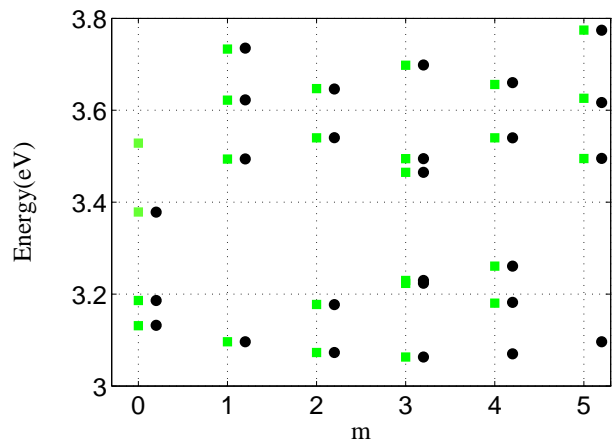}
 \caption{ Energy position of hexagonal modes being in the chosen energy range versus their angular momentum. The circles show the modes belonging to the second ``Brillouin zone'' }
 \label{position}
 \end{figure}

 \begin{figure}[fbp]
 \includegraphics[width=0.99\linewidth]{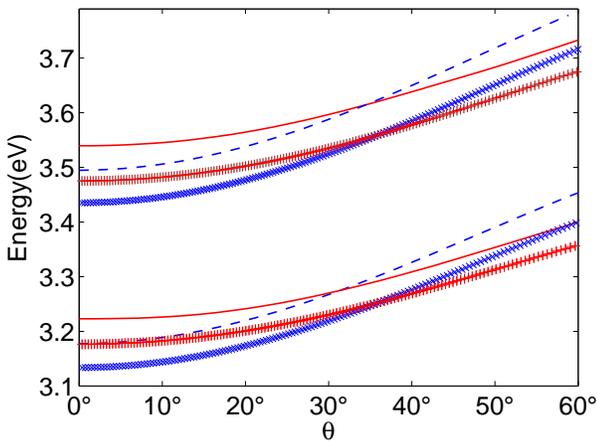}
 \caption{ The lower energy branch show the dispersion of TE and TM modes of circular cross-section wires: TE, $m=8$ ( x symbols, blue) and TM, $m=9$ ( + symbols, red) and the corresponding hexagonal modes: TE, $m=2$ ( dashed blue line) and TM, $m=3$ ( full red line). The same for the upper energy range replacing $m$ by $m+1$.
 }
 \label{dispersion}
 \end{figure}

 \begin{figure}[tbp]
 \includegraphics[width=0.99\linewidth]{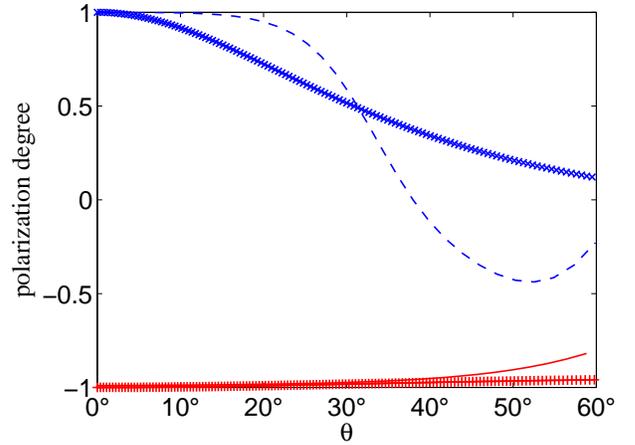}
 \caption{ Polarization degree of the lower energy modes of the Figure~\ref{dispersion}:hexagon modes $m=2$ ( TE, red full line) and  $m=3$ ( TM, blue dashed line) and corresponding cylindrical modes.}
 \label{poldeg}
 \end{figure}

 \begin{figure}[tbp]
 \includegraphics[width=0.99\linewidth]{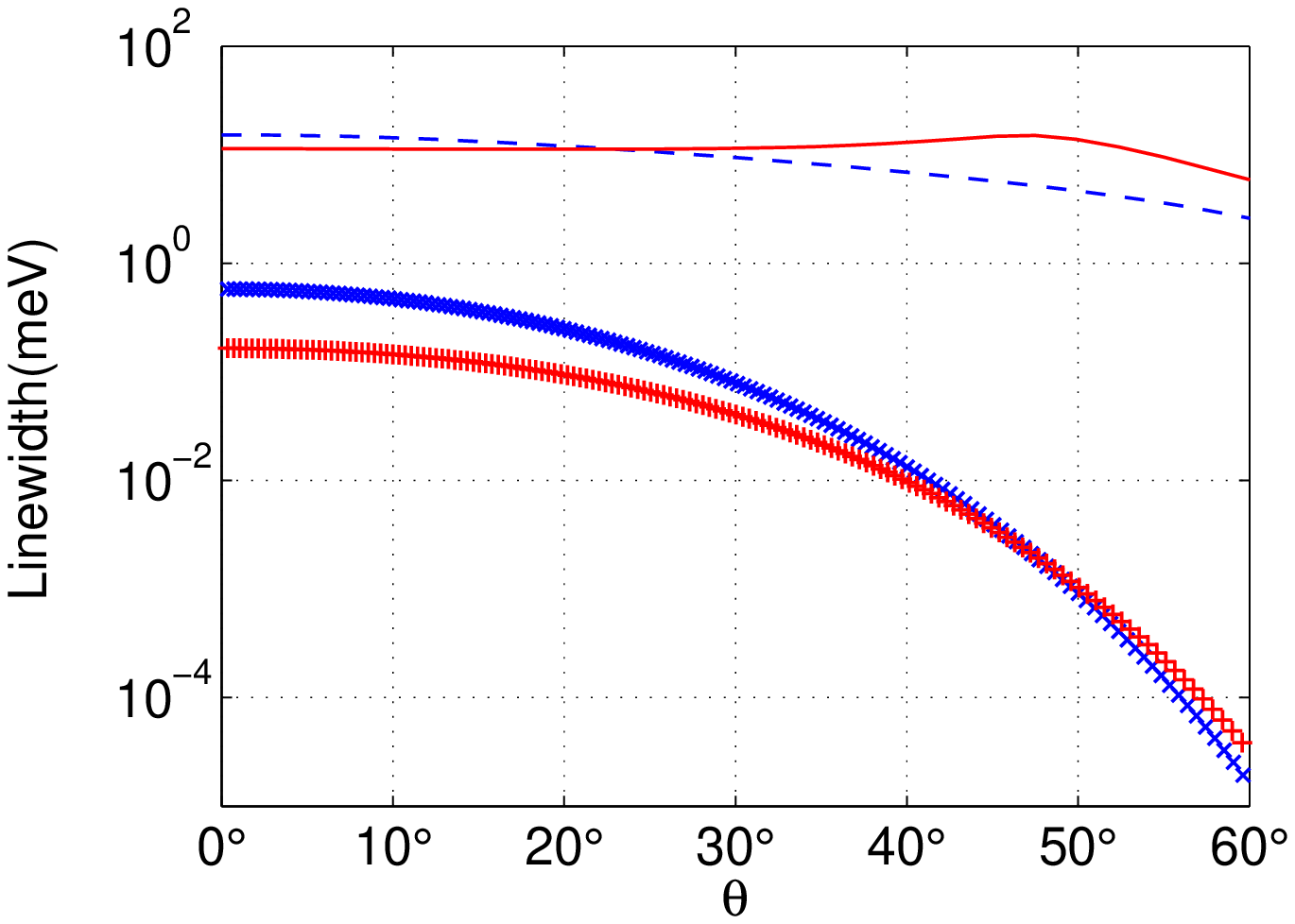}
 \caption{ linewidths of the lower energy modes of the Figure~\ref{dispersion}. The index of refraction is $n=6.25$ and the cross section area $S=0.09\pi \mu m^{2}$
 }
 \label{LW}
 \end{figure}

The matching of inside and outside fields should be realized at any angle ${\rm 0}\le \varphi _{k} {\rm <}2\pi /n$ with corresponding radius $r_{k} =r(\varphi _{k} )$ along the wire boundary. The $r$ and $\varphi$ components in the above expression are given by the equations (\ref{17},\ref{21}) and (\ref{16},\ref{20}). The magnetic field tangential to the NW surface is described in a similar way. In order to keep a convenient matrix description of the problem, we are going to consider boundary conditions only on a finite number of points on the surface, and also to consider a finite number of cylindrical harmonics in the expressions of electric and magnetic fields. In this framework, boundary conditions can be expressed through the matrix equation $\tilde{F}(\omega ,k_{z} )\tilde{X}=0$, where the matrix $\tilde{F}(\omega ,k_{z} )$ is given by:

\begin{equation} \label{26}
\tilde{F}(\omega ,k_{z} )=\left[\begin{array}{ccccc} {F_{11} } & {\cdots } & {F_{1m} } & {\cdots } & {F_{1M} } \\ {\vdots } & {} & {\vdots } & {} & {\vdots } \\ {F_{k1} } & {\ldots } & {F_{km} } & {\ldots } & {F_{kM} } \\ {\vdots } & {} & {\vdots } & {} & {\vdots } \\ {F_{K1} } & {\cdots } & {F_{Km} } & {\cdots } & {F_{KM} } \end{array}\right]
\end{equation}

Here $K$ is the number of points we take on the boundary and $M$ is the number of harmonics we sum up. $F_{km} $ stands for an analog of $F_{m} (\omega ,k_{z} ;r_{k} )$ - the matrices given by the expression (\ref{24}) with tangential $\varphi $-components replaced by (\ref{25}). These sub-matrices describe contribution of a single $m$-harmonic to the boundary condition at the surface point $r_{k} =r(\varphi _{k} )$.
The vector $\tilde{X}$ is now a $4M$-dimensional colon $(X_{1} ,\ldots ,X_{m} ,\ldots ,X_{M} )^{T} $ and $X_{m} $ is defined again like in the case of circular cross-section $X_{m} =(A_{m} ,B_{m} ,C_{m} ,D_{m} )$. The meaning of the matrix (\ref{26}) is that, at any point $r_{k}$ on the NW surface, the same linear combination of cylindrical harmonics allows to verify the boundary conditions at these points. Taking the number of harmonics equal to the number of points on the boundary $K=M$ allows to make the matrix (\ref{26}) square. In this case the eigenmodes $\omega (k_{z} )$ are found as the solutions of the system $\tilde{F}(\omega ,k_{z} )\tilde{X}=0$. Each eigenmode of a $n$-polygonal system contains in addition to the principal harmonic $m$ all other harmonics which add to to it by rule $m+kn$, where $k$ is an non-zero integer. Such eigenmodes, resulting from summation over different harmonics, do not have a well defined angular momentum because of the fact that they are not the eigenstates of the angular momentum operator. Nevertheless, we will associate a number $m$ corresponding to the angular momentum of the principal harmonic to each mode, like in the case of a cylinder. The dispersion of these modes $\omega ({k_z})$ can be found solving $\det (\tilde{F})=0$ or alternatively by a more efficient numerical procedure \cite{Tikhodeev,Gippius}.
 
\section{Results}

In this section we consider the important particular case of the hexagonal cross section and compare the eigenmodes with the ones obtained for a cylinder. The special case of the hexagonal cross section is of strong practical interest since it is realized experimentally by wires made of wurtzite semiconductors such as GaN and ZnO. The comparison with the cylindrical geometry is also particularly relevant since it is a much more simple problem to solve. In practice, modeling of hexagonal NWs is often performed using a cylindrical description.

\begin{figure}[tbp]
 \includegraphics[width=0.99\linewidth]{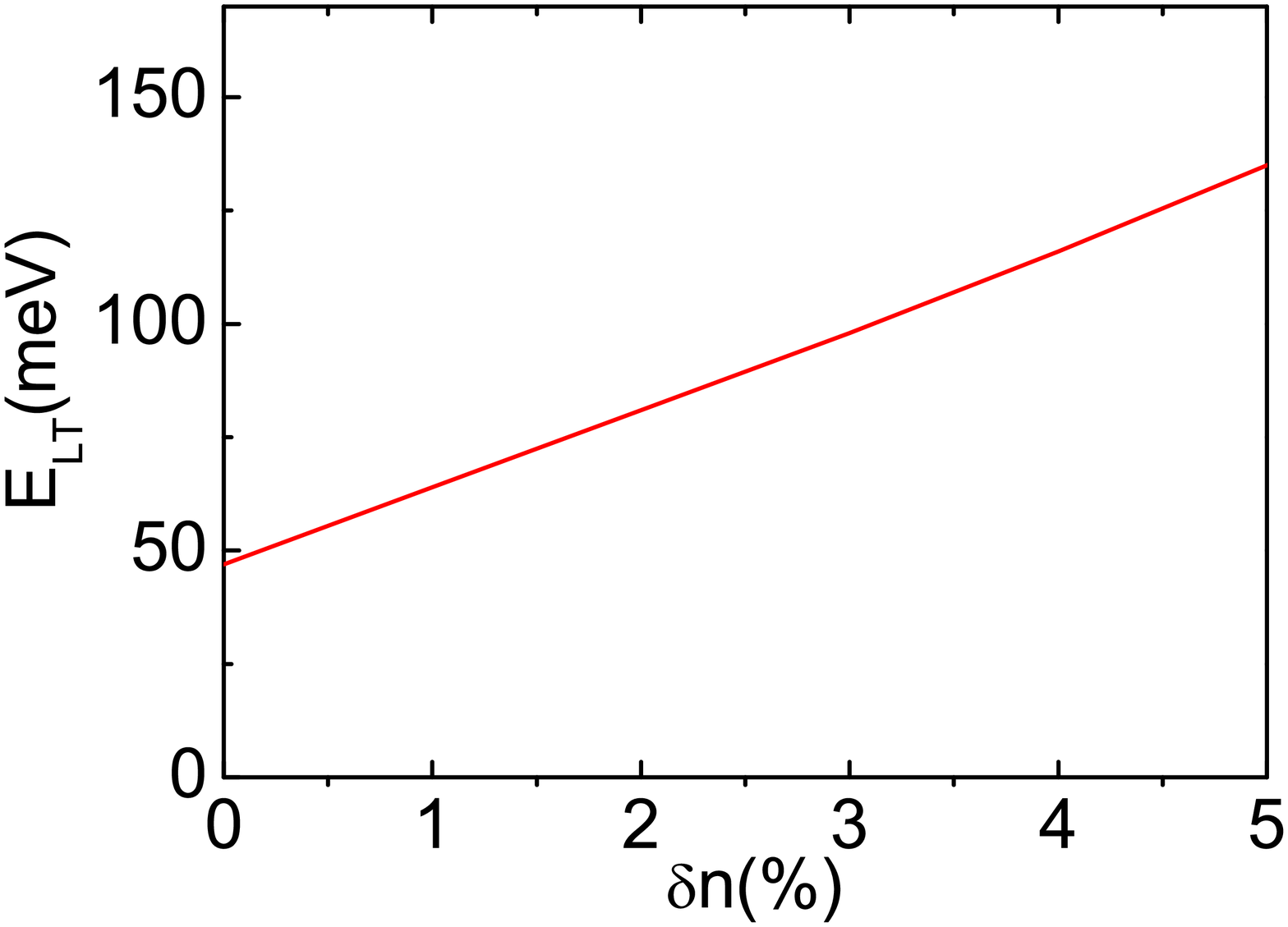}
 \caption{ TE-TM splitting of hexagon modes at zero angle of incident light. The upper two modes of hexagonal NW are shown:TE ( $m=3$) and TM ( $m=4$)}
 \label{Graph1}
 \end{figure}

 \begin{figure}[tbp]
 \includegraphics[width=0.99\linewidth]{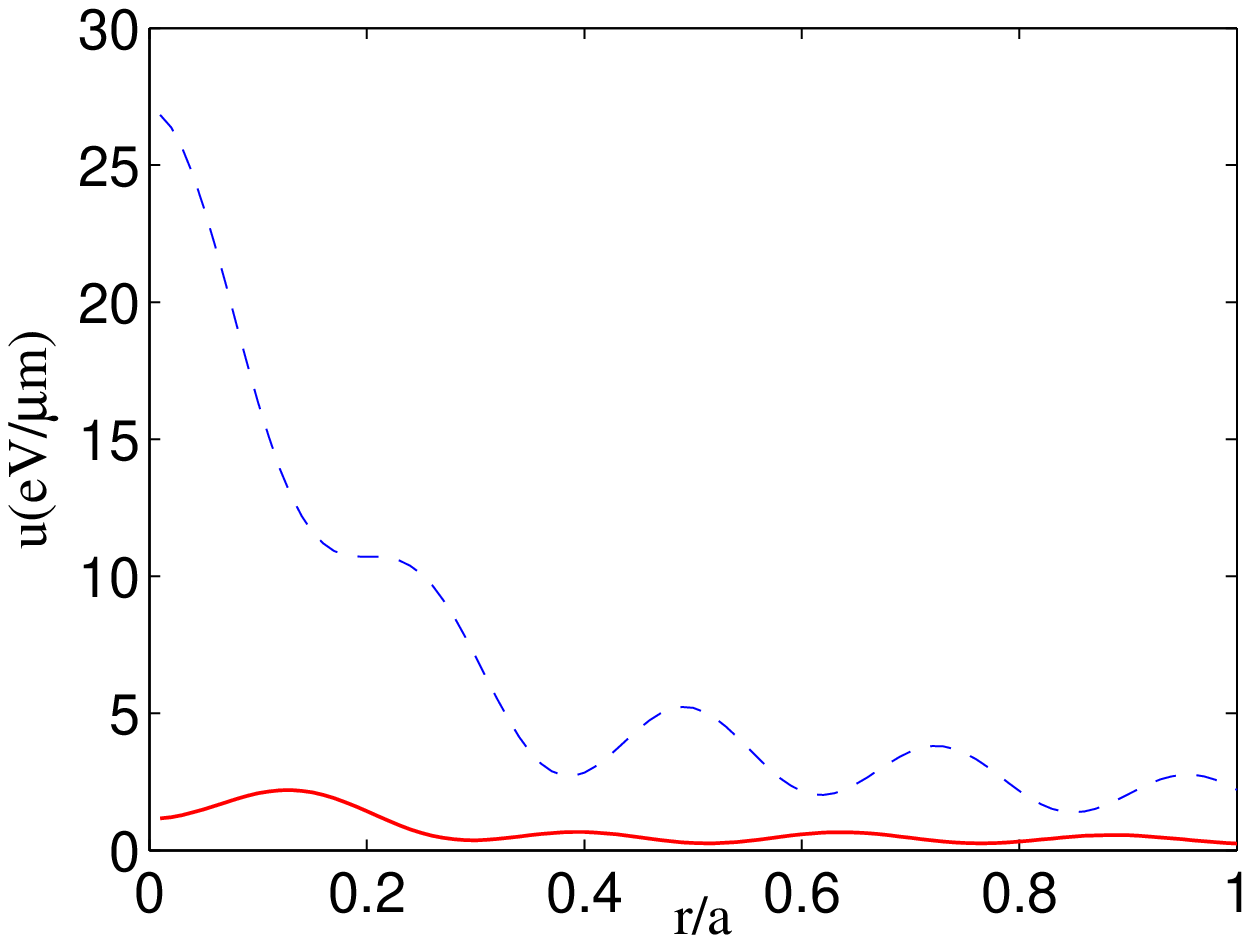}
 \caption{ Energy density radial dependence of the lower pair of hexagon modes:TE ( $m=3$)(full red line) and TM ( $m=4$)(dashed blue line)}
 \label{Fields}
 \end{figure}

We start by deducing empirically, how many cylindrical harmonics should taken into account in order to get a good precision on the energy of the eigenmodes. We consider a NW with a hexagonal cross section and a circumcircle radius of 330 nm. For the Figure~\ref{convergence}, we consider an isotropic dielectric response with an optical index 2.5. We look for the eigen mode $m=0$ in the energy range 3.05-3.25 eV versus the number of cylindrical harmonics $M$ we take into account in the calculation.One can clearly see that the energy of the mode converges for large enough $M$. In all the following calculations, we will use $M=15$.
The energy of the modes is studied in the angular momentum space, the $m$-space (Figure~\ref{position}).  The modes of a hexagonal NW with a circumcircle radius of 330 nm are analyzed for $m=1,...12$, and the energies $m=7,...,12$ are shown with some small shift with respect to the first six harmonics in order to ease the comparison. The numbers of modes and their energies repeat themselves with a periodicity $\Delta m=6$ as expected from the discussion on mode symmetries for a hexagonal system. All relevant physical properties of a hexagonal NW can therefore be deduced looking into the first "Brillouin zone" placed between $m=0$ and $m=5$.
Next, we compare the energy, dispersion, and polarization of hexagonal and circular NWs. Such comparison has been already performed in \cite{Wiersig}, but only from the point of view of the energies of the eigenmodes at $kz=0$. We are focused on the particular energies ( 3.1-3.7 eV, Figure~\ref{dispersion}) corresponding to the energy range in which semiconductor's excitons couple to light, like for example in ZnO NWs \cite{Trichet}. In general, in order to directly compare the properties of the modes in circular and polygonal cross section NWs we have to consider structures having the same cross section area. If the radius of the cylinder is $a$, then the diagonal of corresponding  $n$-side polygon can be found from the formula $d = 2a{(\csc (2\pi /n)2\pi /n)^{1/2}}$.
The Figure~\ref{dispersion} shows the energy dispersions versus the angle of the incident light $\theta  = \arcsin ({{{k_z}} \mathord{\left/ {\vphantom {{{k_z}} k}} \right. \kern-\nulldelimiterspace} k})$ ( Figure~\ref{fig1}(a)) of the modes of two NWs with circular and hexagonal cross-section respectively and having the same area. In both geometries the modes appear in polarization pairs TE and TM in a narrow energy range. The upper mode of the pair is characterized by an angular momentum $m$ and is TM polarized, whereas the lower mode has an angular momentum $m-1$ and is TE polarized.
The modes of the hexagon  appear at higher energies and have a slightly changed dispersion compared to the one of the cylinder. By using slightly different index of refraction for the cylinder and the hexagon, it is possible to match the dispersion of one eigenmode of the two different structures. This shows that in practice the eigenmode of an hexagonal NW can be reasonably described by a simple model assuming a cylindrical geometry.

Both in the cylinder and in the hexagonal wires, the modes are purely TE and TM only at $k_z=0$. The Figure~\ref{poldeg} shows the dependence of the polarization degree $\rho  = ({I_{TE}} - {I_{TM}})/({I_{TE}} + {I_{TM}})$ of the modes versus the incidence angle. In the case of cylindrical wires  $\rho$ can be expressed directly through the coefficients of the external fields as  ${\rho _m} = ({\left| {{D_m}} \right|^2} - {\left| {{C_m}} \right|^2})/({\left| {{C_m}} \right|^2} + {\left| {{D_m}} \right|^2})$. The polarization degree of a TE mode decreases from 1 approaching zero value at higher angles.  The polarization of a TM mode changes very slowly, remaining close to  $-1$ for all  $\theta$. The polarization mixing in hexagonal NWs has a bit different behavior in comparison with the circular geometry case. This difference is most obvious for TE modes. After being almost constant for a wide range of angles,  it starts to decrease significantly near $\theta  \approx 30°$ and becomes even slightly TM polarized between 40 and 60 degrees. On the other hand, the evolution of TM modes is similar to the case of a cylindrical structure.
Such behavior results form the mixing of the hight-$m$ wispering gallery harmonics with low-$m$ ones with larger radial numbers.

The main difference between the hexagon NW and the cylinder is the linewidth of the eigenmodes which is much larger for the hexagon. This is demonstrated on the figure 7 which shows the mode linewidth versus angle of the 4 lower modes shown on the Figure~\ref{dispersion}. The linewidths of the hexagon modes are of the order of 10 meV which is more than an order of magnitude larger than for the cylinder. It is even much larger at higher angles for which the linewidths of the modes of the hexagon remain roughly constant whereas the ones of the cylinder drop by several orders of magnitude. This is due to the presence of corners in a hexagonal NW which are responsible for higher losses resulting in larger linewidths than in a cylindrical wire. This was illustrated in reference \cite{Wiersig} by analysing the mode width dependence versus the rounding of the corners of a hexagon. 

Another important feature of wurtzite materials is their optical birefringence. The optical index along the main c-axis, corresponding to the z-axis of the hexagonal NW differs from the one in the plane. The effect of the birefringence on the longitudinal-transverse splitting $E_{lt}$ (energy splitting of TE and TM polarizations) is shown on the Figure~\ref{Graph1} for $\theta=0$. $E_{LT}$ depends linearly on $\delta n$ and we show dependance on positive birefrigence like it is the case in ZnO wires \cite{Trichet}. Even a small birefringence leads to a significant splitting. It is therefore important to take the anisotropy into account in order to be able to reproduce realistic experimental situations. The radial dependence of the electromagnetic field density  is shown on the Figure~\ref{Fields} for the case of a hexagonal NW.

\section{Conclusions}

In conclusion, we have developed a method which allows to solve Maxwell's equations in NWs of discrete symmetries, and even ones showing an anisotropic dielectric response. This method can be applied to any system having the cross section symmetry of regular polygons. It allows to find the eigenmodes of the structure (whispering gallery modes) labeled by their angular, or pseudo-angular momentum in the case of non-cylindrical structures. The dispersion (dependence of the energy on the wave vector along the wire axis), polarization, linewidth, and radial densities of the modes are calculated for the cases of hexagonal and circular cross sections having the same area respectively. The modes in both cases appear to be quite similar, except from the point of view of the linewidth, which is much larger for hexagonal NWs. We have found some interesting polarization mixing effect with the transformation of TE modes close to $k_z=0$ in modes mainly TM polarized.

The authors would like to thank Maxime Richard, Le Si Dang, and Aur\'elien Trichet for fruitful discussions on this subject.

\end{document}